\documentclass[12pt]{article} 
\usepackage{amssymb}
\usepackage{graphicx}

\newcommand{\ee}{\end{equation}}
\newcommand{\be}{\begin{equation}}
\newcommand{\ba}{\begin{array}}
\newcommand{\ea}{\end{array}}
\newcommand{\bea}{\begin{eqnarray}}
\newcommand{\eea}{\end{eqnarray}}

\title{
Baryogenesis with four-fermion operators in low-scale models}
\author{Thomas Dent\thanks{E-mail: tdent@umich.edu} \\ 
	{\em Michigan Center for Theoretical Physics, Randall Lab.,} \\
	{\em University of Michigan, Ann Arbor, MI 48109-1120}}
\date{October 2001}

\begin{document}
\maketitle
\begin{abstract}
\noindent 
We describe a nonstandard proposal for baryogenesis in models with 
a low (${\mathcal O}(10-100)\,$TeV) fundamental scale. 
The scenario has Standard Model field content and enhanced baryon number 
violating operators deriving from time-dependent fermion localization in 
an extra dimension. The CKM matrix provides sufficient CP violation. The 
major constraints are the low reheating temperature and rate of 
perturbative $B$-violating reactions compared to the total entropy 
created. A sufficient baryon fraction may arise, but the cosmological 
evolution required is likely to be somewhat contrived.  
Based on work in collaboration with D.\,J.\,H.~Chung. 
\end{abstract}

\section{Introduction}
The lower limit on the lifetime of the proton is a severe problem for models
in which the fundamental scale of quantum gravity is low compared to the 
supersymmetric GUT scale $10^{16}\,$GeV \cite{Lykken,AHDD+IAHDD}. 
A baryon number U$(1)$ symmetry cannot be gauged in field theory, so like 
other accidental symmetries it is expected to be violated by effects at the
string scale or by quantum wormholes and virtual black holes 
\cite{BanksDixon_Gilbert}. Such $B$ violation appears at low energies as 
nonrenormalizable operators, for example
\be
\frac{\lambda_{ifgh}}{M_*^2} q_iq_fq_gl_h \label{eq:qqql}
\ee
where $M_*$ is the fundamental scale, $i$, $f$, $g$, $h$ are family labels
and the $\lambda$ are expected to be ${\mathcal O}(1)$ in the absence of 
suppression mechanisms. Then for $M_*$ in the ${\mathcal O}(10-100)\,$TeV 
range, for which collider signals of the fundamental degrees of freedom 
or of large extra dimensions may be observable, $\tau_p$ comes out to be 
under a second. Various solutions have been proposed 
\cite{BenakliD,AH_Schmaltz,Ibanez} all of which have implications 
for the production of an excess of baryons over antibaryons 
in the early Universe, for which $B$ violation is a precondition. 

The overproduction of gravitational Kaluza-Klein modes in such models, in 
which some compactified extra dimensions are orders of magnitude larger
than the fundamental length, gives a severe upper bound on the temperatures 
that can be attained in the early Universe. Even for the maximum number (6) 
of large extra dimensions and the relatively large value $M_* = 100\,$TeV, 
a reheating temperature of a few GeV is the maximum if overclosure of the 
Universe, disruption of the successful predictions of nucleosynthesis, and 
an observationally unacceptable level of background gamma-rays from K-K mode 
decay are to be avoided \cite{BenakliD,Hall+malc+hannestadcosmo}. 
Astrophysical production and decay of such modes also leads to an 
independent lower bound on the fundamental scale 
\cite{CullenBarger_nHannestad}, which is also constrained by the 
non-observation of direct and loop effects in current experiments 
\cite{KKsigs}. 

Any attempt at explaining proton longevity and baryogenesis should operate 
within these constraints. Exact (anomaly-free) discrete or horizontal 
symmetries can be imposed to forbid $B$-violating operators mediating proton 
decay \cite{KraussDGS,BenakliD} while allowing others, through which 
baryogenesis occurs: this approach requires an ``X-boson'', with couplings 
which appear unnaturally small (in contrast to the standard GUT scenario) 
\footnote{Or a charged scalar with $B$-violating couplings which has
a time-dependent v.e.v., in an Affleck-Dine-like model 
\cite{Allahverdi:2001dm}.}. Baryon number can be gauged if the anomaly 
is cancelled by a string theory mechanism, or $B$ violation may be forbidden 
to all orders in perturbation theory by string selection rules, in some 
``intersecting brane'' models \cite{Ibanez}. Note however that in a 
more general class of intersecting brane models \cite{kors_etal}, such 
selection rules do not prevent the four-fermion operators from appearing, 
as discussed above, in which case the fundamental scale cannot be low.

If baryon number is perturbatively exact, nonperturbative processes 
\cite{DavidsonLR} are the only option to create net baryon number. 
It is difficult to see how this proposal can be 
reconciled with cosmological constraints, since any such processes would 
operate at or above the electroweak scale and be enormously suppressed at 
low temperature \footnote{It is unclear whether electroweak baryogenesis 
with low $T_{\rm rh}$ \cite[first reference]{DavidsonLR} actually 
satisfies the bounds, since a hot plasma at temperature $T\gg T_{\rm rh}$
is needed.}.

\begin{figure}[tb] 
\centering{\includegraphics[width=12cm]{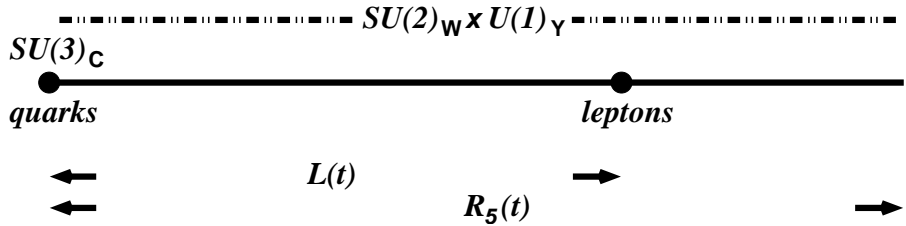}}
\caption{Localization of Standard Model fields in the 5th dimension}
\label{fig:setup}
\end{figure}

We describe a scenario based on a geometrical mechanism for suppressing 4d 
$B$-violating operators, namely localization of fermions in extra dimensions
\cite{AH_Schmaltz}. The simplest implementation is for the SU$(2)\times
{\rm U}(1)$ gauge fields to propagate in an extra dimension ({\it cf.}\/\ 
\cite{DienesDG}), in which the
quark and lepton wavefunctions are peaked about points separated by
a distance $L\sim 30 M_*^{-1}$: see Fig.~\ref{fig:setup}. Then any 
strong $B$-violating operators in the effective 5d theory can only produce 
proton decay proportional to the overlap of the wavefunctions, which is
exponentially small. Alternatively, proton decay by exchange of massive 
modes is suppressed by the Yukawa propagator over the distance 
$L$\footnote{Assuming that no light fermionic modes with $B$-violating
interactions propagate over the bulk.}. Nonperturbative quantum effects 
which may lead to proton decay, for example virtual black holes 
\cite{Adams}, are also exponentially suppressed due to the integration 
over the fifth dimension.\footnote{$B$ violation by electroweak 
sphalerons \cite{KuzminRS}, being an effect {\em energetically}\/ 
suppressed at low temperatures, is not affected by the fermion localization.}

Since the distance $L$ may vary over cosmological time, $B$ violation in the 
$D=4$ field theory may have been unsuppressed ($\lambda\sim 1$) at some 
epoch. Then, if some Standard Model degrees of freedom have number densities 
out of equilibrium, nonzero $n_b$ may be created by the inelastic scattering 
or decay of fermions. A similar scenario was proposed in \cite{Masiero}, but
the authors concentrated on an ``X-boson'' model with somewhat arbitrary 
scalar mass and couplings.
We find that it may be unnecessary to introduce new particles with 
$B$-violating couplings, if the cosmological evolution of number densities
and of the extra-dimensional geometry satisfy certain conditions. 
These appear somewhat special, in that a generic cosmology is unlikely 
to allow our proposal; however, they motivate further study of cosmology 
in low-scale models, since there is no standard picture and few general 
bounds exist on the behaviour before nucleosynthesis.

\section{The model}
The localization of fermions by scalar field profiles has been described
in detail in \cite[and references therein]{AH_Schmaltz}: in essence, 
5d fermions are coupled to a
scalar field profile resulting in a position-dependent effective mass 
$m_5(y)$, where $y$ is the coordinate in the fifth dimension. Then the 
fermion wavefunctions $\psi_{L,R}(y)$ peak about the zeros of $m_5(y)$, 
with the 4d chirality of the localized state determined by the direction
of crossing zero. Localized chiral fermions can also result from orbifold 
projection \cite{GeorgiGH} and coupling to a scalar field odd under the 
orbifold action. Different localized positions can be produced by coupling
to different scalar fields, or by allowing constant 5d fermion masses and 
factors of order 1 in the scalar coupling term. In the approximation that
the scalar field profile is linear near the fermion position $y_f$ a 
Gaussian is obtained
\be
\psi(y) \simeq \mu^{1/2} e^{-\mu^2 (y-y_f)^2},
\ee
where $\mu$ is a parameter of mass dimension 1 which describes the size 
of the scalar v.e.v.: $\Phi \sim \mu^2 y$ near $y=y_f$. Far away from $y_f$,
if the scalar approaches a constant value $\Phi_0$ the wavefunction varies
as $\psi \propto e^{-k\Psi_0|y-y_f|}$, with $k\sim 1$. Then given a 5d 
operator $\lambda^{(5)}M_*^{-3}(QQ)(QL)$, where $(\ )$ denotes the Lorentz- 
and SU$(2)$-invariant sum, the resulting 4d interaction is
\be 
\delta S = \int d^4x\, \frac{\lambda}{M_*^2}(qq)(ql)
\ee
\be {\rm where\ } \lambda \sim 
\lambda^{(5)} \int dy\, \frac{\mu^2}{M^*} e^{-3\mu^2 y^2-\mu^2(y-L)^2} 
\sim \lambda^{(5)} \frac{\mu}{M^*}e^{-3\mu^2L^2/4}
\ee
in the approximation of a linear $\Phi(y)$. For a constant scalar v.e.v.\ 
the suppression goes as $\lambda\propto e^{-k\Phi_0 L}$ with $k\sim 1$.

The presentday value of $L$ required for a sufficiently stable proton is
estimated by comparing with dimension-6 operators induced at the SUSY-GUT 
scale: we require
\be
\lambda^2 M_*^{-4} \simeq M_{\rm GUT}^{-4} \simeq (2\times 10^{16}\,
{\rm GeV})^{-4}
\ee
thus $\lambda \sim e^{-50}$ for $M_*\sim 10-100\,$TeV. Then assuming
$\mu$, $\Phi_0$ to be of the order of $M_*$, up to small numerical factors
we obtain $LM_*\geq 10$ for a linear scalar v.e.v.\ and $LM_*\geq 50$ for a 
constant v.e.v.. This order of magnitude is marginally compatible with 
perturbativity of the bulk SU$(2)$ gauge dynamics, since the $D=5$ gauge
coupling satisfies 
$4\pi R_5 g_{(5)}^{-2}=\alpha_W^{-1} \simeq 31$ and 5d loop corrections
at the fundamental scale are expected to be of order 
$M_*g_{(5)}^{2}/4\pi=\alpha_W M_*R_5$. 

An independent bound on $M_*$ comes from experimental
limits on neutron-antineutron oscillation, which is mediated by a dimension-9 
operator of form $uddudd$. This operator is not suppressed by fermion
localization, thus the coupling strength, of order $M_*^{-5}$, is bounded such 
that $M_* \gtrsim 10^5\,$GeV \cite{BenakliD}. 

\section{Creating $n_b$}
\begin{figure}[tb] 
\centering{\includegraphics[width=14cm]{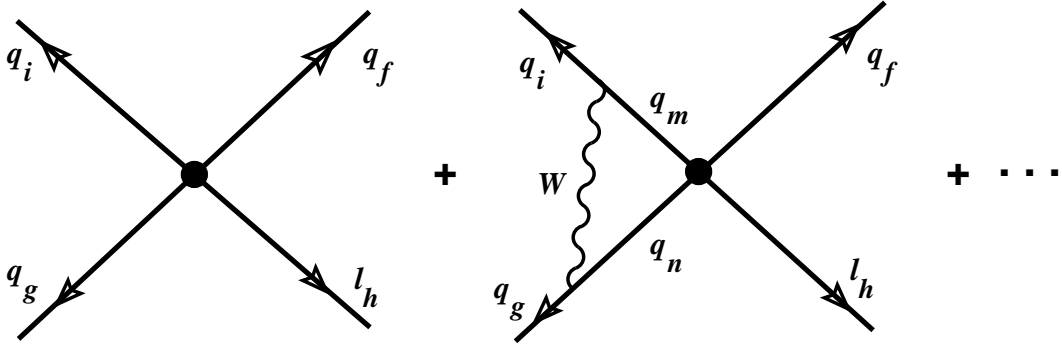}}
\caption{Diagrams interfering to give CP asymmetry: the quartic vertex is
$\lambda/M_*^2$. Other topologies of the $W$ loop are possible.}
\label{fig:diagrams}
\end{figure}

The processes satisfying the Sakharov conditions of $B$, $C$ and $CP$ 
violation and out-of-equilibrium are fermion scattering and decay via the 
dimension-6 operators (\ref{eq:qqql}). $CP$ violation enters by the loop
correction with $W$ exchange, in which CKM matrix elements appear
(Fig.~\ref{fig:diagrams}). 
The asymmetry in cross-section between a process and its $CP$ conjugate 
is proportional to the rephasing invariant combinations of couplings
\be
\eta_{i,f,g,h} \equiv \alpha_W \sum_{nm} 
{\rm Im}\,(\lambda_{nmfh}\lambda^*_{igfh}V_{im}V^*_{ng})
\ee
which are not necessarily correlated with the Jarlskog parameter of the 
SM, and may be of order $10^{-2} - 10^{-1}$. 

Then our procedure is as follows: to place an upper bound on the baryon 
fraction $n_b/s$ we assume that the $\lambda$ are unsuppressed at the 
time of baryogenesis and that we have some fermion number densities 
$n_{f_i}$ well in excess of the thermal equilibrium density: technically, 
we assume kinetic equilibrium (distribution of energy) within each species 
but no chemical equilibrium between species. The upper limit on number 
densities is taken as the equilibrium density at $T=30\,$GeV to avoid 
the possibility of restoring electroweak symmetry. Then the evolution of 
$n_b$ and the entropy density $s$ can be found, given a reasonable
time-dependence of $L(t)$, the 4d scale factor $a(t)$, temperature $T(t)$ 
and the 
extra-dimensional radii $R_i(t)$, $i=5$, \ldots\ as inputs. The effect of 
scattering and decay reactions on the entropy density can be calculated 
directly using the Boltzmann equation \cite{us}, also taking into account 
out-of-equilibrium reactions which 
change number densities but do not change baryon number, for example 
weak decay and inelastic scattering, or annihilation of quarks into gluons.
We find a surprisingly simple result, independent of the details of the 
cosmology: the baryon fraction is bounded above as
\be
\frac{n_b}{s} < 2 \eta \left( \frac{\Gamma_{\rm BV}(t)}{\Gamma_{\rm tot}(t)} 
\right)_{\rm max} \label{eq:gammabound}
\ee
where $\Gamma_{\rm BV}(t)$ is the rate of $B$-violating reactions, 
$\Gamma_{\rm tot}(t)$ is the total rate of reactions changing the number
densities $n_{f_i}$ and the maximum is taken at a time when baryogenesis 
is occurring. The upper bound can be approached if sources of entropy
apart from reactions of the $f_i$ during baryogenesis are small and the 
$B$-violating couplings turn off ({\em i.e.}\/\ $L$ becomes large) soon 
after baryogenesis. Since $B$-violating cross-sections are always suppressed 
as $\alpha_{W} M_*^{-4}$, compared to weak interactions suppressed as 
$\alpha_{W}^2 M_W^{-4}$ or strong scattering cross-sections varying 
as ${\rm max}(m_Q^4,T^4)^{-1} \alpha_{\rm S}^2$, the bound is quite 
restrictive and rules out most possibilities for the $B$-violating reaction. 

The remaining possibilities involve baryogenesis through scattering 
reactions of fermions $f_1$, $f_2$, where the competing $\Delta B=0$ 
reactions are weak decay, annihilation and inelastic scattering. A more  
stringent entropy bound than (\ref{eq:gammabound}) can be derived by 
considering $B$-conserving scattering of $f_1$ with $f_2$, from which 
we find 
\be
\frac{n_b}{s} \leq \frac{\eta}{\alpha_{W}|V_{\rm CKM}|^2}\left( 
\frac{M_{W}}{M_*}\right)^2 \simeq 
4\cdot 10^{-14}(\alpha_{W}|V_{\rm CKM}|^2)^{-1}
\ee
if the inelastic weak scattering is kinematically allowed, where 
$V_{\rm CKM}$ is the relevant matrix element for this reaction. As noted 
above, the weak coupling may change if the radius of the 5th dimension 
varies, but even assuming $\alpha_{W}$ stays small one requires 
$|V_{\rm CKM}|^2$ to be order $10^{-3}$ or smaller even for marginal 
viability. Then the $f_i$ must be such that the weak scattering is 
kinematically suppressed at temperatures above the QCD phase transition
(we do not consider reactions below this temperature, since they are 
complicated by quark confinement). 

There are two candidates for the
reacting fermion species, $us$ (with $u$ out of equilibrium during 
baryogenesis) and $q\nu_\tau$ where $q=c$, $s$, $d$, $u$ (with either 
species out of equilibrium). For the first case, the entropy due to 
{\em self}\/-annihilation reactions can be estimated, with cosmology 
parameterised as $a(t)\propto t^n$, $T\propto a^{-\nu}$, with the result 
that $n_b/s$ is some orders of magnitude below what is required, for any
reasonable value of $n$ and $\nu$ \cite{us}. However, considering the 
self-annihilation reaction of $\nu_\tau$ one finds that if the effective
temperature of the $\nu_\tau$ population  is small enough (below 
approximately $0.4\,$GeV) then the annihilation cross-section
can be arbitrarily small, vanishing (at tree level) in the limit of zero 
temperature and neutrino mass. Thus the rate of entropy creation through
$B$-conserving reactions may be small enough to allow baryon creation to 
compete in this case.

\section{Further discussion}
Our assumptions about the mechanism of baryogenesis, in particular the 
restriction to the Standard Model degrees of freedom, lead to some 
definite conclusions ruling out many possibilities, regardless of
the unknown details of cosmology. The amount of $CP$ violation, a major
problem for baryogenesis in the Standard Model without $B$-violating 
operators and for supersymmetric electroweak baryogenesis if experimental
bounds on soft breaking terms are to be respected, is not an important
constraint for us, even if all $B$-violating couplings are real. The main 
problems for our proposal come from the fact that the species reacting to
create baryon number almost always have strong $\Delta B=0$ reactions that 
compete overwhelmingly with operators suppressed by the fundamental scale. 

We have still to address the questions of how the out-of-equi\-librium 
number density of the particular species involved is to be obtained, and 
whether the 
late evolution of the ``quark-lepton radion'' $L(t)$ can be consistent 
with cosmological and other bounds on scalars associated with the 
geometry of extra dimensions (recall that $L$ should be order 1 at the
time of baryogenesis). It should also be remembered that the estimates of
entropy density that we presented are lower bounds, and any other 
significant sources might alter our conclusions (in a negative direction).

These issues should eventually be addressed within a low-scale
cosmology which includes inflation, reheating, an effective potential
driving the time-dependence of $L(t)$, and a suitable evolution
of the extra-dimensional scale factors. As noted above, there is no 
``standard model'' for cosmology in low-scale theories, primarily because
of the necessity to avoid overproducing light gravitational modes.  
However, extra-dimensional theories have possibilities which might give
grounds for hope, for example if the radii $R_{6-10}$ were smaller at the 
time of baryogenesis, the 4d Planck mass would be smaller and the 
out-of-equilibrium condition might be easier to
satisfy. Also, the early stages of reheating are in general expected to 
produce a nonequilibrium distribution, with thermalization occurring later.
The work described here shows only that if an appropriate 
cosmology can be constructed, baryogenesis through interaction of the SM 
degrees of freedom only is not {\em a priori}\/ ruled out. 

\subsection{Experimental signals}
Even in the absence of a complete model, one can ask whether the scenario
we have described implies testable predictions. Certainly, since we use 
localized fermions, one would expect distinctive collider signatures 
\cite{AH_Schmaltz,Rizzo} in addition to those produced by Kaluza-Klein
modes of gravity and gauge fields \cite{KKsigs}. 
If baryon number is maximally violated in the underlying 
theory then we expect the unsuppressed $n-\bar{n}$ oscillations 
to be detectable, at a rate close to the current experimental limit. 
However, this ``signature'' is rather indirect since it does not occur 
through the same operators as baryogenesis; it would, though, confirm in a 
dramatic fashion the existence of perturbative $B$ violation.
Given a low fundamental scale, the energy density during inflation must 
also be low, which will have definite consequences for cosmology 
\cite{GermanRS}.

\subsection*{Acknowledgements}
The author would like to thank Dan Chung for a
stimulating and rewarding collaboration.
Research supported in part by DOE Grant DE-FG02-95ER40899 Task G.

\end{document}